\documentclass[prb,onecolumn,superscriptaddress,showpacs,preprintnumbers,amsmath,amssymb]{revtex4}

\usepackage{graphicx}% Include figure files
\usepackage{dcolumn}%Align table columns on decimal point
\usepackage{bm}% bold math
\usepackage{epsfig}
\newcommand{\ignore}[1]{}

\begin{document}

\title{An insight into the electronic structure of graphene: from monolayer to multi-layer}

\author{Z. F. Wang}
\affiliation{Hefei National Laboratory for Physical Sciences at
Microscale, University of Science and Technology of China, Hefei,
Anhui 230026, People's Republic of China}

\author{Huaixiu Zheng}
\affiliation{Electrical and Computer Engineering, University of
Alberta, AB T6G 2V4, Canada}

\author{Q. W. Shi} \thanks{Corresponding author. E-mail: phsqw@ustc.edu.cn}
\affiliation{Hefei National Laboratory for Physical Sciences at
Microscale, University of Science and Technology of China, Hefei,
Anhui 230026, People's Republic of China}

\author{Jie Chen}
\affiliation{Electrical and Computer Engineering, University of
Alberta, AB T6G 2V4, Canada}

\author{Jinlong Yang}
\affiliation{Hefei National Laboratory for Physical Sciences at
Microscale, University of Science and Technology of China, Hefei,
Anhui 230026, People's Republic of China}

\author{J. G. Hou}
\affiliation{Hefei National Laboratory for Physical Sciences at
Microscale, University of Science and Technology of China, Hefei,
Anhui 230026, People's Republic of China}

\date{\today}

\begin{abstract}
In this paper, we analytically investigate the electronic
structure of Bernal stacking (AB stacking) graphene evolving
from monolayer (a zero-gap semiconductor with a linear Dirac-like
spectrum around the Fermi energy) to multi-layer (semi-metal bulk graphite).
We firstly derive a real space analytical expression for the free
Green's function (propagator) of multi-layer graphene based on the
effective-mass approximation. The simulation results exhibit highly spatial anisotropy
with three-fold rotational symmetry. By combining with the STM measurement
of $d^2I/dV^2$ (the second derivative of current), we also provide a clear
high-throughput and non-destructive method to identify graphene layers.
Such a method is lacking in the emerging graphene research.
\ignore{
We discover that the electronic structure of monolayer graphene differs significantly from
that of bilayer upto multilayer graphene,
but the electronic structure of bilayer is similiar to that of multilayer.}

\end{abstract}


\maketitle

\section{INTRODUCTION}
The successful synthesis of monolayer graphite (graphene)
has attracted enormous research interest over the past years. \cite{1}
Researchers can also experimentally manipulate few layer within multi-layer
graphene samples and observe their quasi two-dimensional behavior. \cite{2,3,4,5,6,7}
The conduction band of graphene is well described by the  tight-binding model,
which includes the $\pi$ orbitals that is perpendicular to the plane each
carbon atom locates. This model describes a semi-metal, with zero density
of states at the Fermi energy. Here, the Fermi surface is reduced
to two inequivalent K-points located at the corners
of the hexagonal Brillouin Zone. The low-energy excitations
with momenta in the vicinity of any Fermi
points have a linear dispersion. They can be described by
a continuous model, which reduces to the Dirac equation
in two dimensions. These observations have been validated experimentally \cite{1}.

Recent research attention has turned to study multi-layer
graphene \cite{2,3,4,5,6,7}, because its electronic proprieties are
quite different from those of the monolayer graphene. Two-dimensional
(2D) multi-layer graphene is an intermediate crystals between bulk graphite
and a single graphene plane. Relatively weak inter-layer coupling within
multilayer graphene inherits some properties from monolayer graphenes.
The special geometry of the Bernal stacking (A-B stacking) between graphene
layers results in low-energy states mainly reside around B site.
In contrast to corresponding three-dimensional (3D) bulk structures,
electrons in a multi-layer graphene are confined along one crystallographic
direction, which offers its unique electronic characteristics. From the Hall
effect measurements of a multi-layer graphene, the results show that multi-layer
graphene behaves like multi-carriers (coexistence of electrons and holes)
semi-metallic systems. \cite{8,9} Moreover, carriers can switch from electron
to hole transport by alteranting gate voltage. This phenomenon makes multi-layer
graphene a remarkable platform for studying mesoscopic transport in
low-dimensional materials. It is also worth noting that multi-layer graphene
systems may be of interest for buidling nanoscale devices because we can easily
change their electronic properties by using conventional lithographic techniques.

To date, few studies attempting to uncover the unique
electronic properties of multi-layer graphene in an analytical fashion.
Moreover, the physical properties of multi-layer graphene are
closely related to its propagator. In this paper, we first develop
an analytical formula of the free Green's function for electrons in
multi-layer graphene (Bernal stacking) in real space by using the
effective-mass approximation. We then calculate the LDOS of multi-layer
graphene and isotropic function by using Green's function explicitly
with increasing number of graphene layers. Our numerical results show
that monolayer graphene is dramatically different from other multi-layer
graphene, and bilayer graphene capture the main characters of the
multi-layer graphene. Currently only AFM and Raman spectrum \cite{1,10,11} are used to
detect the layers of graphene. In this paper, we present an effective
method to detect graphene layers based on the STM.
According to our derivative of LDOS with energy
(proportional to the second derivative of current $d^2I/dV^2$),
we can quickly, clearly and nondistructively identify the graphene
layers of finite multi-layer graphene based on STM images.

\section{Tight-BINDING DESCRIPTION}

\begin{figure}
\begin{center}
\epsfig{figure=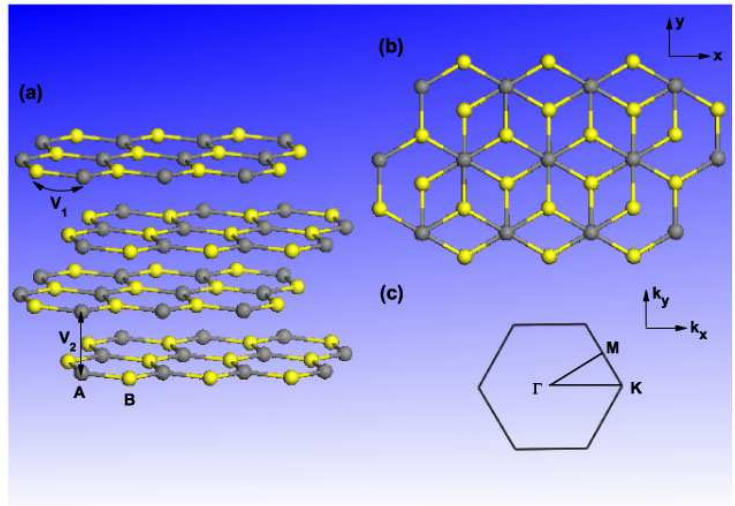,height=8.0cm}
\caption{(color online)
The crystal structure of four AB stacked graphene layers.
Type-A atoms (gray) have direct neighbors in the
adjacent atomic layers while type-B atoms (yellow) are above hollow sites.
(a) Side view with the corresponding tight-binding parameters, (b)
Top view. (c) The first Brillouin zone of a finite
number of graphene layers with the labels of
special symmetric points (Here, we choose four-layer graphene as an example to
illustrate our design). }
\end{center}
\end{figure}

Figure 1  shows schematically the crystal structure of four
AB-stacked graphene layers and the corresponding first Brillouin
zone with the labels of the symmetric points. It was shown in the
reference [12] that the interplanar spacing parameter $a$ and lattice
spacing $d$ in two-layer graphene are almost identical to that in
bulk graphite. We, therefore, choose the bulk value $a=1.42{\AA}$
and $d=3.35{\AA}$ in this paper. In order to study how the electronic
structure of graphene changes from monolayer into multi-layer, we
construct the tight-binding Hamiltonian for an arbitrary number
of graphene layers. We extend our previous studies of monolayer
graphene \cite{13} and of bilayer graphene \cite{14} to multi-layer one.
Starting from two-layer graphene, type A and B carbon atoms are
inequivalent: type-A carbon atom has a direct neighbor
in its adjacent layer while type-B carbon atom does not and
locates on the hollow sites. Since the physical properties of
graphene are determined by $\pi$ bands near the Dirac point,
only the contribution from the $\pi$ band is considered in this paper.
In our tight-binding model, we limit our calculation to the
nearest interaction carbon atoms on the graphene's $p_z$ orbitals.
These interactions include interactions between the nearest A-B carbon
atoms within a plane, interaction between nearest A-A carbon atoms of
two nearest-neighbor planes. The tight-binding Hamiltonian of the
multi-layer graphene is:

\begin{equation}
\begin{split}
H=V_1\sum_{l=1}^L\sum_{{\langle}ij\rangle}{|i^l\rangle}{\langle}j^l|
+V_2\sum_{{\langle}ll'\rangle}\sum_{{\langle}ij\rangle}{|i^l\rangle}{\langle}j^{l'}|.
\end{split}
\end{equation}

\noindent where ${\langle}\textbf{r}|i^l\rangle$ is the wavefunction
at site $i$ on layer $l$. $L$ is the number of graphene layers.
$V_1$ is the nearest hopping parameter
within the layer and $V_2$ is the nearest hopping parameter between
two layers. We select $V_1=-3.0 eV$ and $V_2=0.4 eV$ in our calculation.

The Bloch orbits for two nonequivalent sites, A and B, on layer $l$ are written as
${|\textbf{k}_A^l\rangle}=\frac{1}{\sqrt{N}}\sum\limits_{l_A}
e^{i\textbf{k}\cdot\textbf{r}_{l_A}}|\textbf{r}_{l_A}\rangle
\mspace{5mu}$,
${|\textbf{k}_B^l\rangle}=\frac{1}{\sqrt{N}}\sum\limits_{l_B}
e^{i\textbf{k}\cdot\textbf{r}_{l_B}}|\textbf{r}_{l_B}\rangle$.
\noindent The summation is over all sites A and B on layer $l$.
$\textbf{r}_{l_A}$ and $\textbf{r}_{l_B}$ denote the real-space coordinates
of A and B sites, respectively. Here, N is the number of unit cells
in the crystal. The tight-binding Hamiltonian can be obtained from the following
matrix elements:

\begin{eqnarray}
{\langle \textbf{k}_A^l|}H{|\textbf{k}_A^l\rangle}&=&{\langle \textbf{k}_B^l|}H{|\textbf{k}_B^l\rangle}=0 ,\nonumber\\
{\langle \textbf{k}_A^l|}H{|\textbf{k}_B^l\rangle}&=&\left\{\begin{array}{cc}
V_1\mu^* & (l=odd) \\
V_1\mu   & (l=even)\end{array}, \right.\nonumber\\
{\langle \textbf{k}_B^l|}H{|\textbf{k}_A^l\rangle}&=&\left\{\begin{array}{cc}
V_1\mu & (l=odd) \\
V_1\mu^*   & (l=even)\end{array}, \right.\nonumber\\
{\langle \textbf{k}_A^l|}H{|\textbf{k}_A^{l+1}\rangle}&=&{\langle \textbf{k}_A^{l+1}|}H{|\textbf{k}_A^l\rangle}=V_2.
\end{eqnarray}

\noindent with
\begin{eqnarray}
\begin{split}
\mu=exp(ik_ya)+exp[i(-\frac{\sqrt{3}ak_x}{2}-\frac{ak_y}{2})]
+exp[i(\frac{\sqrt{3}ak_x}{2}-\frac{ak_y}{2})]. \nonumber
\end{split}
\end{eqnarray}

\noindent
In these caculations, we neglect all overlap integrals because of the relatively
large separation between the carbon atoms.  Furthermore, we
are only interested in small energy electronic structure
around the K point. With the expressions in Eq (2), we can construct
the tight-binding Hamiltonian for an arbitrary number of layers

\begin{eqnarray}
H=\left(\begin{array}{ccccc}
0 & V_1\mu^* & V_2 & 0 & \cdots\\
V_1\mu & 0 & 0 & 0 & \cdots\\
V_2 & 0 & 0 & V_1\mu & \cdots\\
0 & 0 & V_1\mu^* & 0 & \cdots\\
\vdots & \vdots & \vdots & \vdots & \vdots\
\end{array}\right).
\end{eqnarray}

The eigenfunction for  multi-layer graphene system
is now given by
\begin{eqnarray}
|\textbf{k}\rangle=\sum_{l=1}^L\phi_A^l|\textbf{k}_A^l\rangle+\sum_{l=1}^L\phi_B^l|\textbf{k}_B^l\rangle.
\end{eqnarray}
The hopping between sublattices of different layer lead to the coupled Harper equations,

\begin{eqnarray}
\left\{\begin{array}{cc}
\varepsilon\phi_A^l=&V_2\phi_A^{l-1}+V_1\mu^*\phi_B^l+V_2\phi_A^{l+1}\\
\varepsilon\phi_B^l=&V_1\mu\phi_A^l\end{array} \right. \mspace{10mu} (l=odd).
\end{eqnarray}

and
\begin{eqnarray}
\left\{\begin{array}{cc}
\varepsilon\phi_A^l=&V_2\phi_A^{l-1}+V_1\mu\phi_B^l+V_2\phi_A^{l+1}\\
\varepsilon\phi_B^l=&V_1\mu^*\phi_A^l\end{array} \right. \mspace{10mu} (l=even).
\end{eqnarray}

\noindent where $l=1,2,\cdots, L$ denotes the layer number.
The open boundary condition in the direction perpendicular
to the 2D graphene plane is
$\phi_A^0=\phi_A^{L+1}=\phi_B^0=\phi_B^{L+1}=0$.
According to the coupled Harper equations Eq.(5) and Eq.(6), the wave
functions of the eigenstates and the corresponding energy
spectrum can be obtained in the following analytical form.

\begin{eqnarray}
\varepsilon_m^s(\textbf{k})=V_2cos(\frac{m\pi}{L+1})+s\sqrt{V_2^2cos^2(\frac{m\pi}{L+1})
+V_1^2\mu\mu^*}
\end{eqnarray}
\begin{eqnarray}
\phi_A^l(m,s,\textbf{k})&=&D(m,s,\textbf{k})sin(\frac{ml\pi}{L+1}),\nonumber\\
\phi_B^l(m,s,\textbf{k})&=&
\left\{\begin{array}{cccc}
D(m,s,\textbf{k})\frac{V_1\mu}{\varepsilon_m^s(\textbf{k})}sin(\frac{ml\pi}{L+1})
\mspace{20mu} (l=odd)\\
D(m,s,\textbf{k})\frac{V_1\mu^*}{\varepsilon_m^s(\textbf{k})}sin(\frac{ml\pi}{L+1})
\mspace{20mu} (l=even)
\end{array}, \right.
\end{eqnarray}

\noindent where
$D(m,s,\textbf{k})=\{\sum\limits_{l=1}^{L}[1+\frac{V_1^2\mu\mu^*}{\varepsilon_m^s(\textbf{k})^2}]
sin^2(\frac{ml\pi}{L+1})\}^{-1/2}$ is the normalized factor $(m=1,2,\cdots,L)$ and $s=\pm 1$.

\section{REAL-SPACE GREEN'S FUNCTION}
Based on the definition of retarded Green's function, we obtain the reciprocal space
Green's function for multi-layer graphene as:

\begin{eqnarray}
G^{ret}_{0,l_\mu l'_\nu}(m,\textbf{k},E)=\sum\limits_{s=\pm1}\frac{\phi_\mu^l(m,s,\textbf{k}) \phi_\nu^{l'*}(m,s,\textbf{k})}
{E+i\eta-\varepsilon_m^s(\textbf{k})},
\end{eqnarray}

\noindent where $\eta$ is an infinitesimal quantity, $\mu$ and $\nu$ label A or B.
By taking the Fourier transform of
$G^{ret}_{0,l_\mu l'_\nu}(m,\textbf{k},E)$
in the first Brillouin zone (1BZ), we can obtain the exact expression
of real space multi-layer graphene Green's function,

\begin{eqnarray}
G^{ret}_{0}(\textbf{r}_{l_\mu},\textbf{r}'_{l'_\nu},E)=
\sum\limits_{m=1}^{L}
\int_{1BZ}d\textbf{k}G^{ret}_{0,l_\mu l'_\nu}(m,\textbf{k},E)
e^{i\textbf{k}\cdot(\textbf{r}_{l_\mu}-\textbf{r}'_{l'_\nu})}
\end{eqnarray}

Based on the effective-mass approximation, the integral is carried out
around six corners in the first Brillouin zone within the low-energy
region, which can form two $360^\circ$ integrals around two inequivalent
Dirac points. This method has been well used to study the electronic structure of
monolayer and bilayer graphene in our previous works. \cite{13,14}
The real space multi-layer Green's function can be obtained as following:

\begin{eqnarray}
%AA
G^{ret}_{0}(\textbf{r}_{l_A},\textbf{r}'_{l'_A},E)&=&
cos[\textbf{K}^1\cdot(\textbf{r}_{l_A}-\textbf{r}'_{l'_A})]F_1(|\textbf{r}_{l_A}-\textbf{r}'_{l'_A}|,E),\nonumber\\
%BB
G^{ret}_{0}(\textbf{r}_{l_B},\textbf{r}'_{l'_B},E)&=&
\left\{\begin{array}{cccccc}
cos[\textbf{K}^1\cdot(\textbf{r}_{l_B}-\textbf{r}'_{l'_B})]F_2(|\textbf{r}_{l_B}-\textbf{r}'_{l'_B}|,E)
\mspace{20mu} (l=odd, l'=odd \mspace{10mu} or \mspace{10mu} l=even, l'=even)\\
cos[\textbf{K}^1\cdot(\textbf{r}_{l_B}-\textbf{r}'_{l'_B})-2\alpha_{r_{l_B},r'_{l'_B}}]F_3(|\textbf{r}_{l_B}-\textbf{r}'_{l'_B}|,E)
\mspace{20mu} (l=odd, l'=even)\\
cos[\textbf{K}^1\cdot(\textbf{r}_{l_B}-\textbf{r}'_{l'_B})+2\alpha_{r_{l_B},r'_{l'_B}}]F_3(|\textbf{r}_{l_B}-\textbf{r}'_{l'_B}|,E)
\mspace{20mu} (l=even, l'=odd)\\
\end{array}, \right.\nonumber\\
%AB
G^{ret}_{0}(\textbf{r}_{l_A},\textbf{r}'_{l'_B},E)&=&
\left\{\begin{array}{cccc}
sin[\textbf{K}^1\cdot(\textbf{r}_{l_A}-\textbf{r}'_{l'_B})+\alpha_{r_{l_A},r'_{l'_B}}]F_4(|\textbf{r}_{l_A}-\textbf{r}'_{l'_B}|,E)
\mspace{20mu} (l'=odd)\\
sin[\textbf{K}^1\cdot(\textbf{r}_{l_A}-\textbf{r}'_{l'_B})-\alpha_{r_{l_A},r'_{l'_B}}]F_4(|\textbf{r}_{l_A}-\textbf{r}'_{l'_B}|,E)
\mspace{20mu} (l'=even)\\
\end{array}, \right.\nonumber\\
%BA
G^{ret}_{0}(\textbf{r}_{l_B},\textbf{r}'_{l'_A},E)&=&
\left\{\begin{array}{cccc}
sin[\textbf{K}^1\cdot(\textbf{r}_{l_B}-\textbf{r}'_{l'_A})-\alpha_{r_{l_B},r'_{l'_A}}]F_4(|\textbf{r}_{l_B}-\textbf{r}'_{l'_A}|,E)
\mspace{20mu} (l=odd)\\
sin[\textbf{K}^1\cdot(\textbf{r}_{l_B}-\textbf{r}'_{l'_A})+\alpha_{r_{l_B},r'_{l'_A}}]F_4(|\textbf{r}_{l_B}-\textbf{r}'_{l'_A}|,E)
\mspace{20mu} (l=even)\\
\end{array}. \right.\nonumber\\
\end{eqnarray}

where
\begin{eqnarray}
F_1(|\textbf{r}_{l_\mu}-\textbf{r}'_{l'_\nu}|,E)&=&\sum\limits_{s=\pm1}\sum\limits_{m=1}^{L}
\frac{2\tilde{S}}{\pi}\int^{k_c}_0dk
\frac{k\tilde{D}^2(m,s,\textbf{k})sin(\frac{ml\pi}{L+1})sin(\frac{ml'\pi}{L+1})J_0(k|\textbf{r}_{l_\mu}-\textbf{r}'_{l'_\nu})|)}
{E+i\eta-\{V_2cos(\frac{m\pi}{L+1})+s[V_2^2cos^2(\frac{m\pi}{L+1})+(rk)^2]^{1/2}\}},\nonumber\\
F_2(|\textbf{r}_{l_\mu}-\textbf{r}'_{l'_\nu}|,E)&=&\sum\limits_{s=\pm1}\sum\limits_{m=1}^{L}
\frac{2\tilde{S}\gamma^2}{\pi}\int^{k_c}_0dk
\frac{\frac{k^3\tilde{D}^2(m,s,\textbf{k})sin(\frac{ml\pi}{L+1})sin(\frac{ml'\pi}{L+1})
J_0(k|\textbf{r}_{l_\mu}-\textbf{r}'_{l'_\nu}|)}{\{V_2cos(\frac{m\pi}{L+1})+s[V_2^2cos^2(\frac{m\pi}{L+1})+(rk)^2]^{1/2}\}^2}}
{E+i\eta-\{V_2cos(\frac{m\pi}{L+1})+s[V_2^2cos^2(\frac{m\pi}{L+1})+(rk)^2]^{1/2}\}},\nonumber\\
F_3(|\textbf{r}_{l_\mu}-\textbf{r}'_{l'_\nu}|,E)&=&\sum\limits_{s=\pm1}\sum\limits_{m=1}^{L}
-\frac{2\tilde{S}\gamma^2}{\pi}\int^{k_c}_0dk
\frac{\frac{k^3\tilde{D}^2(m,s,\textbf{k})sin(\frac{ml\pi}{L+1})sin(\frac{ml'\pi}{L+1})
J_2(k|\textbf{r}_{l_\mu}-\textbf{r}'_{l'_\nu}|)}{\{V_2cos(\frac{m\pi}{L+1})+s[V_2^2cos^2(\frac{m\pi}{L+1})+(rk)^2]^{1/2}\}^2}}
{E+i\eta-\{V_2cos(\frac{m\pi}{L+1})+s[V_2^2cos^2(\frac{m\pi}{L+1})+(rk)^2]^{1/2}\}},\nonumber\\
F_4(|\textbf{r}_{l_\mu}-\textbf{r}'_{l'_\nu}|,E)&=&\sum\limits_{s=\pm1}\sum\limits_{m=1}^{L}
\frac{2\tilde{S}\gamma}{\pi}\int^{k_c}_0dk
\frac{\frac{k^2\tilde{D}^2(m,s,\textbf{k})sin(\frac{ml\pi}{L+1})sin(\frac{ml'\pi}{L+1})
J_1(k|\textbf{r}_{l_\mu}-\textbf{r}'_{l'_\nu}|)}{V_2cos(\frac{m\pi}{L+1})+s[V_2^2cos^2(\frac{m\pi}{L+1})+(rk)^2]^{1/2}}}
{E+i\eta-\{V_2cos(\frac{m\pi}{L+1})+s[V_2^2cos^2(\frac{m\pi}{L+1})+(rk)^2]^{1/2}\}}.\nonumber\\
\end{eqnarray}

\begin{figure}[htpb]
\begin{center}
\epsfig{figure=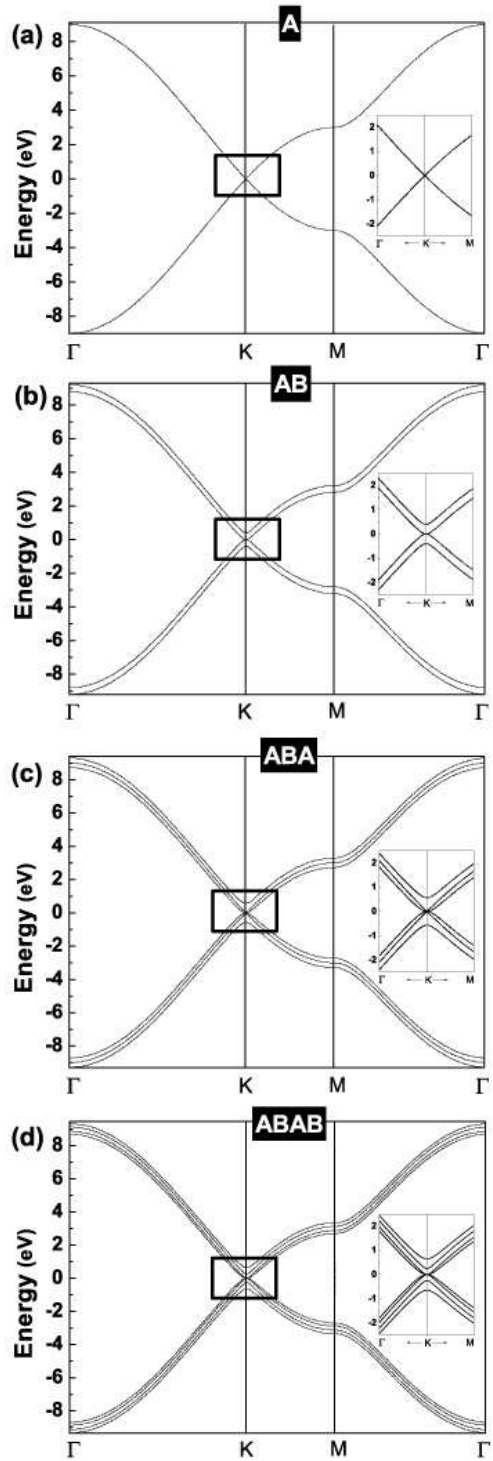,height=22cm}
\caption{The band structure of (a) monolayer, (b) bilayer,
(c) trilayer and (d) 4-layer bernal stacking graphene along
$\Gamma K M \Gamma$. The inset is an enlargement of the region
indicated by the rectangular around the K point.}
\end{center}
\end{figure}

\noindent $\tilde{D}(m,s,\textbf{k})$
is the normalized factor under the effective mass approximation. It can be
obtained from the expression of $D(m,s,\textbf{k})$ by substituting
$V_1^2\mu\mu^*$ with $(\gamma k)^2$. Here, $\gamma=3aV_1/2$.
$\alpha_{r_\mu,r'_\nu}$ is the angle between
$\textbf{r}_\mu-\textbf{r}'_\nu$ and x axis. $\textbf{K}^1=(4\pi/3\sqrt{3}a,0)$
and $J_n$ is the type-I $n$-order Bessel function.
$\tilde{S}=3\sqrt{3}a^2/2$ is the area of the
unit cell in real space and $k_c$ is the cutoff wave vector.
From Eq.(11) and Eq.(12), it is clear that the real space
Green's function of multi-layer graphene is constructed by multiplying
two terms: spatially anisotropic and spatially isotropic terms.
The spatially anisotropic (three-fold rotational
symmetry) can be represented by sine and cosine functions
which can be determined from two nonequivalent sets of the Dirac points.
The second term including $F_1$, $F_2$, $F_3$, $F_4$ is
spatially isotropic in real space and depends only on the distance
between two sites. For L=1 and L=2, the expression of the Green's function
obtain in Eq.(11) and Eq.(12) can directly reduce to our previous results
for monolayer and bilayer graphene. \cite{13,14}

\section{NUMERICAL RESULTS}

\subsection {Band structure}

As an example of applying the dispersion relation obtained in Eq.(7),
we compare the energy band structure of bernal stacking multi-layer
graphene involves from monolayer to 4-layer, the corresponding
energy bands along the $\Gamma K M \Gamma$ lines in the first
Brillouin zone are shown in Fig.2(a) $\sim$ 2(d), respectively.
In all the calculations, the Fermi energy is set to $E_f=0$.
The following are our observations:

(i) For a  monolayer graphene structure as shown in Fig.2(a), the two bands cross
at the K point leads to the fact that a monolayer graphene is
a zero-gap metal. Around the K point the spectrum is
linear as shown in the inset and it is given by $\varepsilon_1^s=s\gamma k$.
Here, k is wave vector from the Dirac K point.
Comparing this expression with the relativistic energy
expression $\varepsilon=\sqrt{m^2c^4+p^2c^2}$, one can see that the dispersion
relation of monolayer graphene mimics a system of
relativistic Dirac particles with zero rest mass and an effective
speed of light $c=10^6 m/s$, which is 300 times smaller than
the speed of light in vacuum.

(ii) For a bilayer graphene structure as shown in Fig.2(b), the number of
levels is doubled. The spectrum is clearly no longer linear around
the K point, but is parabolic. $\varepsilon_1^s=V_2/2+s\sqrt{(V_2/2)^2+(\gamma k)^2}$,
$\varepsilon_2^s=-V_2/2+s\sqrt{(V_2/2)^2+(\gamma k)^2}$.
If moving away from the K point, the spectrum becomes linear again.
In the present calculation by only considering the nearest interaction between layers,
we observe that the bands just touch at the K point. However, a more detailed
investigation based on first principle method
shows a small overlap and an interaction leading to anticrossings between the
conduction and valence bands. \cite{15}

(iii) For a trilayer graphene structure as shown in Fig.2(c), there are four
bands cross at the K point around zero energy, two of them are linear.
$\varepsilon_1^s=\sqrt{2}V_2/2+s\sqrt{(\sqrt{2}V_2/2)^2+(\gamma k)^2}$,
$\varepsilon_2^s=s\gamma k$,
$\varepsilon_3^s=-\sqrt{2}V_2/2+s\sqrt{(\sqrt{2}V_2/2)^2+(\gamma k)^2}$.
It is clear that the band structure around the K point becomes more complex
with increasing number of graphene layers than monolyaer and bilayer graphenes.
For instance, the number of layers around the Fermi energy at the K point is
doubled in comparison with monolayer and bilayer graphenes.
The band diagram can be understood as the combination of band structure
in monolayer graphene and in bilayer graphene.

(iv) For a 4-layer graphene, shown in Fig.2(d), four subbands
cross exist at the K point. From Eq.(7), in the frame work of the nearest
tight-binding model a general number of the subband cross at the K
point can be deduced. For the number of layers (L), there are 2L subband.
L of them cross at the K point if L is even, while L+1 subbands cross at
the K point if L is odd. The additional one subband, existing only when
$cos(\frac{m\pi}{L+1})=0$ (i.e. $m=\frac{L+1}{2}$), is linear.

\subsection{Surface Green's Function}

\begin{figure}
\begin{center}
\epsfig{figure=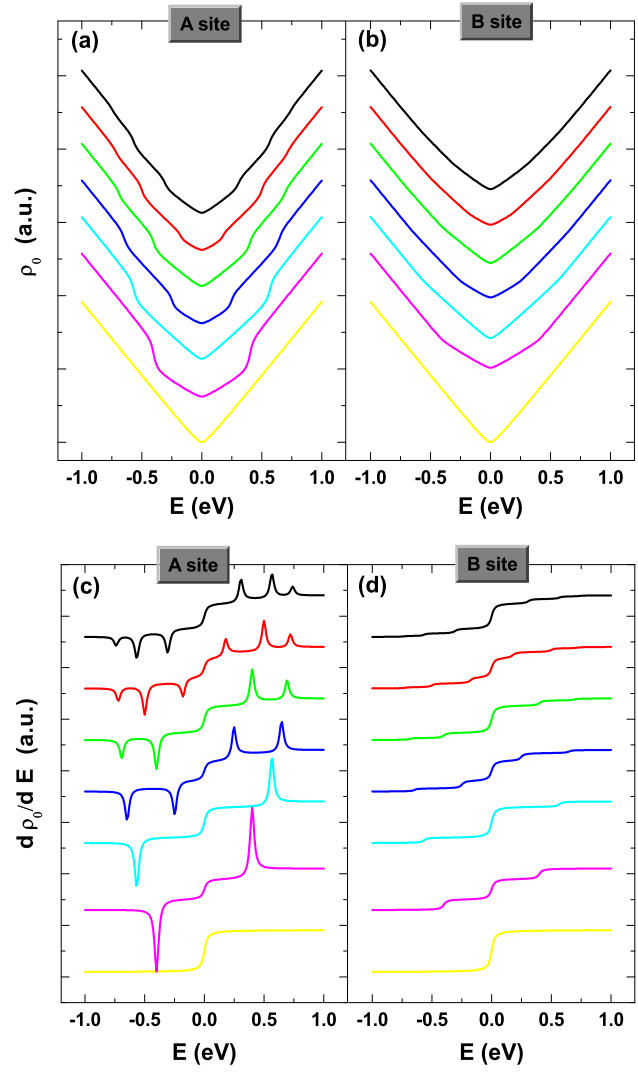,height=18cm}
\caption{(color online)
(a) and (b) are the LDOS of A and B site at the top layer with
different graphene layers (from down to up L=1 $\sim$ 7),
(c) and (d) are the corresponding derivative of LDOS with energy.
To clearly see the difference between them, we shift each curve in (a) $\sim$ (d) upwards.}
\end{center}
\end{figure}

\begin{figure}
\begin{center}
\epsfig{figure=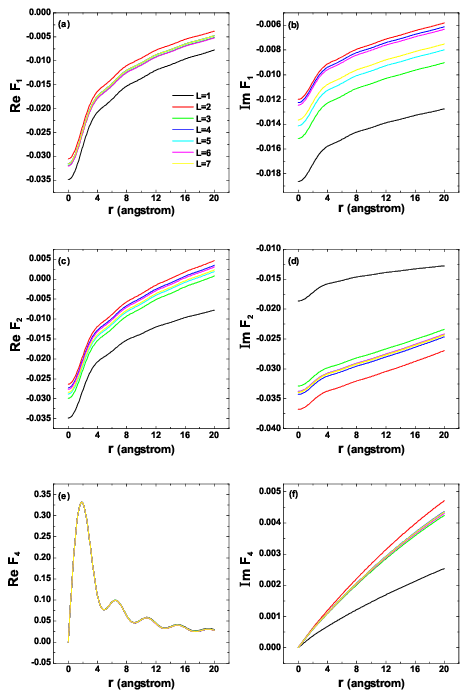,height=20cm}
\caption{(color online) Spatially isotropic function
$F_1$, $F_2$ and $F_4$ of the top layer with different
graphene layers (L=1 $\sim$ 7). Here we set $E=0.1 eV$.}
\end{center}
\end{figure}

In this section, we study the behavior of the top layer
Green's function with the changing number of graphene layers (L) in detail.
From Eq.(11) and Eq.(12), the corresponding real space Green's function
of the top layer $(l=l'=1)$ are
\begin{eqnarray}
G^{ret}_{0}(\textbf{r}_{1_A},\textbf{r}'_{1_A},E)&=&
cos[\textbf{K}^1\cdot(\textbf{r}_{1_A}-\textbf{r}'_{1_A})]F_1(|\textbf{r}_{1_A}-\textbf{r}'_{1_A}|,E)\nonumber\\
G^{ret}_{0}(\textbf{r}_{1_B},\textbf{r}'_{1_B},E)&=&
cos[\textbf{K}^1\cdot(\textbf{r}_{1_B}-\textbf{r}'_{1_B})]F_2(|\textbf{r}_{1_B}-\textbf{r}'_{1_B}|,E)\nonumber\\
G^{ret}_{0}(\textbf{r}_{1_A},\textbf{r}'_{1_B},E)&=&
sin[\textbf{K}^1\cdot(\textbf{r}_{1_A}-\textbf{r}'_{1_B})+\alpha_{r_{1_A},r'_{1_B}}]F_4(|\textbf{r}_{1_A}-\textbf{r}'_{1_B}|,E)\nonumber\\
G^{ret}_{0}(\textbf{r}_{1_B},\textbf{r}'_{1_A},E)&=&
sin[\textbf{K}^1\cdot(\textbf{r}_{1_B}-\textbf{r}'_{1_A})-\alpha_{r_{1_B},r'_{1_A}}]F_4(|\textbf{r}_{1_B}-\textbf{r}'_{1_A}|,E)\nonumber\\
\end{eqnarray}

From previous studies, we find that in the low bias voltage, sites A and B
are equivalent for monolayer graphene. They form honeycomb lattice
in the STM image. While for a bilayer graphene, sites A and B are inequivalent,
and the LDOS is residual at site B and they form triangular lattice. Therefore,
one direct approach to find the difference among multi-layer graphene
is to study the LDOS at sites A and B by changing the number of stacked layers.
From Eq.(13), LDOS at sites A and B at the top layer can be expressed as:

\begin{eqnarray}
\rho_0(\textbf{r}_{1_A},E)
&=&-\frac{1}{\pi}\mbox{Im}G^{ret}_{0}(\textbf{r}_{1_A},\textbf{r}_{1_A},E)\nonumber\\
&=&-\frac{1}{\pi}\mbox{Im}F_1(0,E)\nonumber\\
\rho_0(\textbf{r}_{1_B},E)
&=&-\frac{1}{\pi}\mbox{Im}G^{ret}_{0}(\textbf{r}_{1_B},\textbf{r}_{1_B},E)\nonumber\\
&=&-\frac{1}{\pi}\mbox{Im}F_2(0,E)\nonumber\\
\end{eqnarray}

Figure 3(a) and (b) present our numerical results of LDOS at
sites A and B with different number of graphene layers.
The LDOS for L=1 is linear, which is clearly different from
other curves. Starting from L=2, increasing the number of stacked
layer, the LDOS just shows a slight deviation from its bilayer value
for both A and B sites. This character means that monolayer
graphene is a special case in the family of multi-layer graphene.
The relatively weak inter-layer coupling within multi-layer graphene
not only inherits some properties from monolayer graphenes, but also
solidify some proprieties between different stacked layers. Detailed
studies also find that LDOS at A site converge more slowly than B
site with increasing the graphene layers (Twenty layers are needed
for the LDOS at A site to converge to its bulk value, while just ten
layers needed for B site). LDOS at A site fluctuate more dramatically
than that at B site in the process of convergence.

Since the first synthesis of monolayer graphene,
except AFM and Raman spectrum \cite{1,10,11}, few method can be
used to exactly detect the number of graphene layers. In our work,
based on STM second derivative map, we develop a new method to measure
graphene layers, which is lacking in the emerging graphene research.
From Eq. (14), the derivation of the LDOS with respect to the energy
($d\rho_0/dE$) is proportional to the second derivation of current
($d^2I/dV^2$) in the STM measurement, which corresponds to the
excitation of vibrational modes \cite{16}. The results are shown in Fig.3(c) and (d).
The curves are symmetry to the positive and negative energy. So for
simplicity, we just consider the positive energy section in the following
discussion. For site A, monolayer graphene has no peak, bilayer has one
and trilayer has one as well. General relation for the peak number equals
$Int(L/2)$ (the function $Int$ round off the variable
to an integer). For L is even number, L and L+1 has the same number of peaks,
but the peaks for L+1 move towards the high energy direction relative to the L.
With increasing the number of stacked layers, the intensity of the
peak become much smaller. For L=2, the peak height is about 0.16 $eV^{-2}$.
While for L=14, the highest peak reduce to 0.06 $eV^{-2}$. Additionally, the
position of small peak is hard to determine when L become large than 14.
Therefore, in the real experiment, only the sample has a few layers
(less than 14 layers) may be detected by considering the sensitivity of the
STM measurement. For site B, we observe no peak exists, but just step
curve. With increasing stack layer, the step become not so clearly anymore.

In the above discussion, we mainly focus on the LDOS of the top
layer for the multi-layer graphene. However, to study the propagating behavior
of electrons in the top layer of multi-layer graphene, we also need to know
the properties of the space isotropic function $F_1$, $F_2$, and $F_4$ in
Eq.(12). The numerical results are shown in Fig.4. For different stacked
layer, the curves are almost parallel to each other as shown in Fig.4(a)$\sim$4(d).
Only the curve for L=1 is clearly differs from the others. In Fig.4(e),
the curves overlap with each other and can not be clearly separated.
In Fig.4(f), all curves start from the same point, but with different
slope. The slop of other curves changes slightly except for the curve L=1.
These results indicate that (i) the isotropic function for
monolayer graphene is dramatically different from multi-layer graphene;
(ii) from L=2 onwards, the isotropic function changes little with
increasing the stacked layers. This is consistent with the results
obtained from the information of the LDOS study that monolayer graphene
is a special case in multi-layer graphene. Starting from bilayer, the electronic strcuture of
graphene changes little when layers increase.

\section{CONCLUSION}
In summary, an analytical form of the real space Green's function
(propagator) of multi-layer graphene is firstly constructed by using
the effective-mass approximation. The Green's function demonstrates
elegant spatial anisotropy with three-fold symmetry. Our numerical
results show that the monolayer graphene is dramatically different from
multi-layer graphene, and it is a special species in the family of
graphite. In addition, we provides a new feasible method to identify
the graphene layers by measuring the $d^2I/dV^2$ in STM experiment, or
the predicted features based on our simulated results can be used to verified
STM measurements. Since multi-layer graphene is described by
using the simple non-interactive tight-binding scheme, we are currently
investigating how the Coulomb interaction plays a role in electronic
structure of multi-layer graphene.

\section*{ACKNOWLEDGMENTS}
This work is partially supported by the National Natural
Science Foundation of China under Grants No. 10574119,
10674121, 20533030, 20303015, and 50121202 by National
Key Basic Research Program under Grant No. 2006CB0L1200,
by the USTC-HP HPC project, and by the SCCAS and Shanghai
Supercomputer Center. Work at NTU is supported in part by
COE-SUG Grant No. M58070001. J.C. would like to acknowledge
the funding support from the Discovery program of Natural
Sciences and Engineering Research
Council of Canada (No. 245680).


\begin{references}
\bibitem{1}
K. S. Novoselov, A. K. Geim, S. V. Morozov, D. Jiang, M. I.
Katsnelson, I. V. Grigorieva, S. V. Dubonos, and A. A. Firsov,
Nature \textbf{438},197 (2005).

\bibitem{2}
Yuanbo Zhang, Yan-Wen Tan, Horst L. Stormer and Philip Kim,
Nature \textbf{438},201 (2005).

\bibitem{3}
A. C. Ferrari, J. C. Meyer, V. Scardaci, C. Casiraghi, M. Lazzeri,
F. Mauri, S. Piscanec, D. Jiang, K. S. Novoselov, S. Roth, and A. K. Geim,
Phys. Rev. Lett. \textbf{97},187401 (2006).

\bibitem{4}
Claire Berger, Zhimin Song, Xuebin Li, Xiaosong Wu, Nate Brown, C\'{e}cile Naud,
Didier Mayou, Tianbo Li, Joanna Hass, Alexei N. Marchenkov, Edward H. Conrad,
Phillip N. First, Walt A. de Heer,
Science \textbf{312},1191 (2006).

\bibitem{5}
M.L. Sadowski, G. Martinez, and M. Potemski, C. Berger and W.A. de Heer,
cond-mat/0605739.

\bibitem{6}
J. Scott Bunch, Yuval Yaish, Markus Brink, Kirill Bolotin, and Paul L. McEuen,
Nano Lett \textbf{5},287 (2005).

\bibitem{7}
Davy Graf,. Fran¢¬coise Molitor, Klaus Ensslin, Christoph Stampfer,
Alain Jungen, Christofer Hierold and Ludger Wirtz,
cond-mat/0607562.

\bibitem{8}
K. S. Novoselov, A. K. Geim, S. V. Morozov, D. Jiang, Y. Zhang,
S. V. Dubonos, I. V. Grigorieva, and A. A. Firsov,
Science \textbf{306}, 666 (2004).

\bibitem{9}
Claire Berger,Zhimin Song, Tianbo Li, Xuebin Li, Asmerom Y. Ogbazghi, Rui Feng,
Zhenting Dai, Alexei N. Marchenkov, Edward H. Conrad, Phillip N. First, and
Walt A. de Heer,
J. Phys. Chem. B \textbf{108}, 19 912 (2004).

\bibitem{10}
A. C. Ferrari,J. C. Meyer, V. Scardaci, C. Casiraghi, M. Lazzeri,
F. Mauri, S. Piscanec, D. Jiang, K. S. Novoselov,  S. Roth, and A. K. Geim,
Phys. Rev. Lett. \textbf{97},187401 (2006).

\bibitem{11}
A. Gupta, G. Chen, P. Joshi, S. Tadigadapa, and P.C. Eklund,
Nano Lett \textbf{6},2667 (2006).

\bibitem{12}
S. B. Trickey, F. M¨¹ller-Plathe, G. H. F. Diercksen, and J. C. Boettger,
Phys. Rev. B \textbf{45}, 4460 (1992).

\bibitem{13}
Z. F. Wang, Ruoxi Xiang, Q. W. Shi, Jinlong Yang, Xiaoping Wang,
J. G. Hou, and Jie Chen,
Phys. Rev. B \textbf{74}, 125417 (2006).

\bibitem{14}
Z. F. Wang, Qunxiang Li, Haibin Su, Xiaoping Wang, Q. W. Shi, Jie Chen,
Jinlong Yang and J. G. Hou,
Phys. Rev. B \textbf{75}, 085424 (2006).

\bibitem{15}
Sylvain Latil and Luc Henrard
Phys. Rev. Lett. \textbf{97},036803 (2006).

\bibitem{16}
C. Didiot, Y. Fagot-Revurat, S. Pons, B. Kierren, C. Chatelain, and D. Malterre,
Phys. Rev. B \textbf{74}, 081404(R) (2006).
\end{references}
\end{document}